\documentclass[12pt,a4paper]{article}
\usepackage[utf8]{inputenc}
\usepackage[T1]{fontenc}
\usepackage{amsmath,bbm,cite}
\usepackage{amsfonts}
\usepackage{amssymb,comment}
\usepackage{graphicx}
\usepackage{transparent}
\usepackage{notoccite}
\usepackage{ulem}
\usepackage[left=2cm, right=2cm]{geometry}
\usepackage{todonotes}
\usepackage{xcolor}
\usepackage{physics}
\usepackage{bbold}
\usepackage{ntheorem}
\usepackage{setspace} 
\usepackage[overlay,absolute]{textpos}
\usepackage{stackengine}
\newcommand\xrowht[2][0]{\addstackgap[.5\dimexpr#2\relax]{\vphantom{#1}}}
\usepackage{hyperref}

\newcommand{\be}{\begin{equation}}
	\newcommand{\ee}{\end{equation}}
\newcommand{\bea}{\begin{eqnarray}}
	\newcommand{\eea}{\end{eqnarray}}

\definecolor{AH}{HTML}{d91f05}
\definecolor{BF}{HTML}{f903d7}

\newcommand\PlaceText[3]{%
\begin{textblock*}{10in}(#1,#2)
#3
\end{textblock*}
}%

\begin{document}
\sloppy

\PlaceText{165mm}{15mm}{27 March 2024}
\begin{center}
{\LARGE\bf The Boundary Proposal}

\end{center} 

\vspace*{0.8cm}
\thispagestyle{empty}

\centerline{\large 
Bjoern Hassfeld and
Arthur Hebecker}
\vspace{0.5cm}
 
\begin{center}
{\it Institute for Theoretical Physics, Heidelberg University,}
\\
{\it Philosophenweg 19, 69120 Heidelberg, Germany}\\[0.5ex]  
\end{center} 
\vspace{.25cm}
\centerline{\small\textit{E-Mail:} \href{mailto:hassfeld@wisc.edu}{hassfeld@wisc.edu}, \href{mailto:a.hebecker@thphys.uni-heidelberg.de}{a.hebecker@thphys.uni-heidelberg.de}}

\vspace*{.5cm}
\begin{abstract}\normalsize
One of the leading ideas for the beginning of the Universe is the Hartle-Hawking `No-Boundary Proposal.' Since the Cobordism Conjecture claims that any spacetime allows for a dynamical boundary, we suggest that one may equally well consider a `Boundary Proposal'. Specifically, the corresponding euclidean instanton is a sphere with two holes around north and south pole cut out. Analogously to the Hartle-Hawking proposal, the sphere is then cut in two at the equator and half of it is dropped. The equator is glued to an expanding Lorentzian de Sitter space, implementing a beginning of the Universe with a spacelike spherical boundary at its earliest moment. This process is in principle on equal footing with the one based on the no-boundary instanton. In fact, if the Linde-Vilenkin sign choice is used, this `Boundary' creation process may even dominate. An intriguing implication arises if tensionless end-of-the-world branes, as familiar from type-IIA or M-theory, are available: Analogously to the Boundary Proposal, one may then be able to create a compact, flat torus universe from nothing, without any exponential suppression or enhancement factors.

\vspace*{.4cm}
\noindent
\end{abstract}
\vspace{10pt}

\newpage
 
	\sloppy
\title{TBD}
\date{}
\author{}

\begin{spacing}{0.85}
\setcounter{tocdepth}{2}
    \tableofcontents
\end{spacing}

\section{Introduction}
The idea of vacuum creation from nothing has a long history: Vilenkin, Hartle-Hawking  and Linde  have proposed that the creation rate of de Sitter universes can be obtained by studying appropriate gravitational instantons \cite{Vilenkin:1982de, Vilenkin:1983xq, hartle1983wave, Linde:1983mx, Vilenkin:1984wp} (see \cite{Linde:1984ir, Linde:1990flp} for reviews). In fact, the relevant instanton is simply euclidean dS space, i.e.~a sphere. Alternatively, one may start from the Lorentzian path integral \cite{Halliwell:1988ik, Feldbrugge:2017kzv}. The relevant instantons then appear in the analytic continuation as those saddle points which give the dominant contribution (see \cite{Lehners:2023yrj, Maldacena:2024uhs} for a recent review and critical discussions). It was furthermore observed that, if end-of-the-world (ETW) branes exist,
universes with boundaries may be created from nothing. The corresponding instanton, first introduced in \cite{Hawking:1998bn,Turok:1998he} and further studied in \cite{Garriga:1998ri,Garriga:1998tm,Blanco-Pillado:2011fcm,Hassfeld:2023tid}, has the topology of a ball with a spherical ETW brane as its boundary. The geometry of the creation process arises by cutting this ball in two and gluing one of its halves to an expanding Lorentzian universe with a cylindrical boundary.

In this work, we propose a new type of creation process in which a boundary-free, spherical, de Sitter universe appears from nothing thanks to an initial, space-like boundary, provided by an ETW brane. The corresponding instanton may be thought of as sphere with two holes cut out. This results in two spherical boundaries. The size of the holes follows from the ETW-brane tension and the de Sitter cosmological constant. To interpret this instanton as part of the complete geometry, it is cut in two and one half is glued to an expanding, boundary-free spherical universe, precisely as in the Hartle-Hawking case. The resulting spacetime has one boundary, provided by one of the two holes discussed above (for an illustration see Table~\ref{tab_comparison}). This boundary may be viewed as an initial surface for the universe. Hence the name `Boundary Proposal', as opposed to `No Boundary Proposal', appears fitting.

Importantly, the instanton action approaches zero in the limit of vanishing ETW brane tension. In this situation, the two holes discussed above eat up almost the whole sphere, such that only a thin strip around the equator is left. As a result, there is no tunneling suppression and our one-boundary creation process dominates over the no-boundary process for the Linde-Vilenkin sign choice in the instanton exponent.

\section{A new instanton for the creation of universes}
\subsection{Instanton geometries and actions}\label{sect_instantons}
We consider the euclidean action
\begin{align}
    \mathcal{S}=-\frac{M_P^2}{2}\int_{\mathcal{M}} d^4x \sqrt{g}\left(\mathcal{R}-2\Lambda\right)-M_P^2\int_{\partial \mathcal{M}} d^3x \sqrt{h}\left(\mathcal{K}-T\right)\,,\label{action}
\end{align}
and take the metric to be of FLRW form:
\begin{align}
    ds^2=dt^2+a(t)^2d\Omega_3^2\,.\label{metric}
\end{align}
Here, $\mathcal{R}$ is the Ricci scalar and $\Lambda = 3/\ell_{dS}^2 > 0$ is the cosmological constant. If a boundary $\partial\mathcal{M}$ with extrinsic curvature $\mathcal{K}$ is present, it is the location of an end-of-the-world (ETW) brane with tension $T$. Away from the boundary, the classical solution to the resulting Einstein equations is simply dS space. Three possible types of instantons and the corresponding creation processes are illustrated in Table~\ref{tab_comparison}:

The first corresponds to the  No-Boundary Proposal \cite{Vilenkin:1982de,Vilenkin:1983xq, Vilenkin:1984wp,Linde:1983mx, hartle1983wave}, with the instanton being simply a euclidean $S^4$. It can be interpreted along the lines of the `bounce solutions' of Coleman and de Luccia \cite{coleman1980gravitational}:
A small $S^3$ grows, reaches maximal radius at the equator of the $S^4$, and shrinks again to zero size.  Following the Coleman procedure, one can continue to Lorentzian signature at the turning point. Both the instanton and the resulting creation process are shown in the left column of the table.
\begin{table}[!ht]
\begin{tabular}{c|c|c|c}
     & No-Boundary (nb) & Bubble of Something (bos) & Boundary (b) \\
     \hline
      && &
     \\[-0.5em]
     &\includegraphics[width=0.25\textwidth]{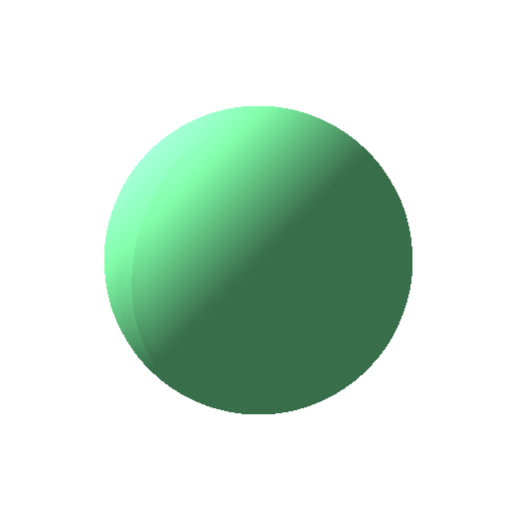}&  \includegraphics[width=0.25\textwidth]{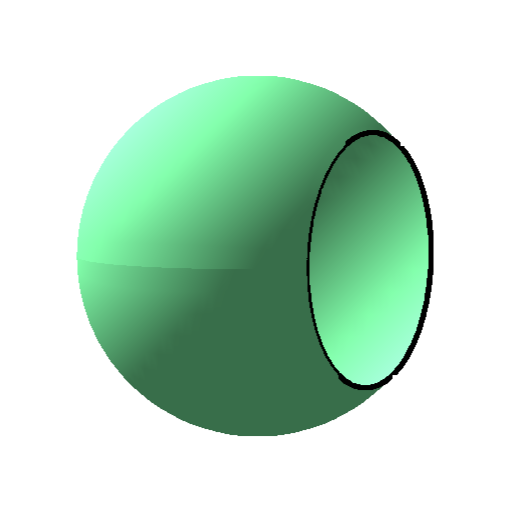} & \includegraphics[width=0.25\textwidth]{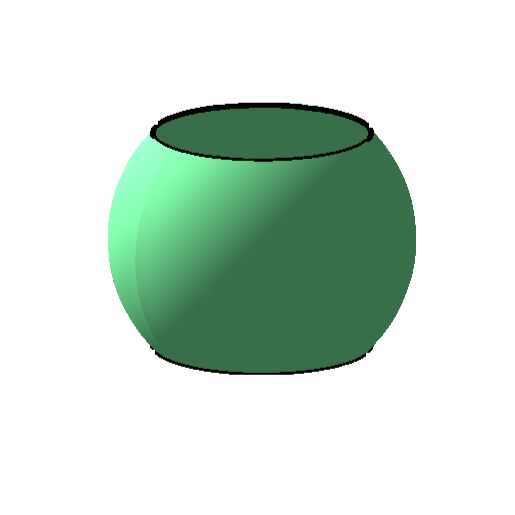}\\
        & & &
     \\[-0.5em]
     \hline
     &&&
     \\[-0.5em]
     &\includegraphics[width=0.25\textwidth]{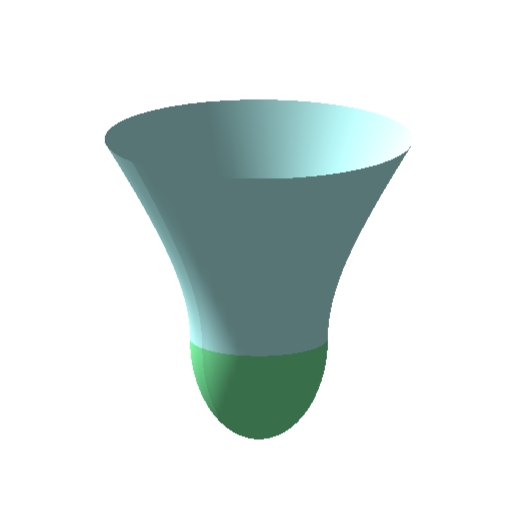} & \includegraphics[width=0.25\textwidth]{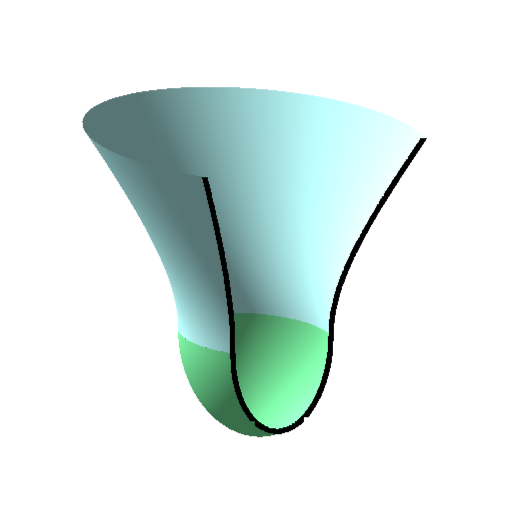} & \includegraphics[width=0.25\textwidth]{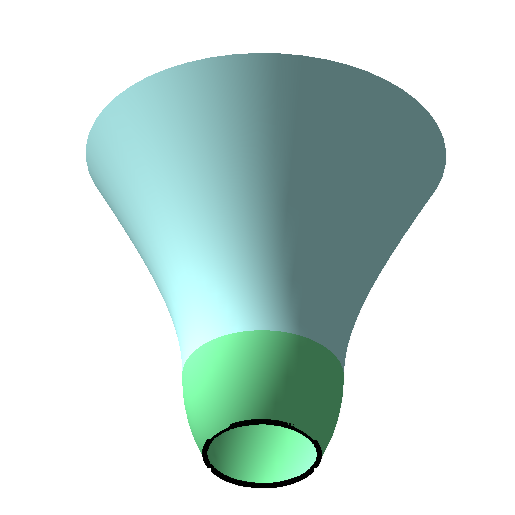}\\
     &&&
     \\[-0.5em]
     \hline\xrowht{30pt}
     $\mathcal{S}=\;$ & $-8\pi^2M_P^2\ell_{dS}^2$ & $-4\pi^2M_P^2\ell_{dS}^2\left(1-\frac{T\ell_{dS}}{\sqrt{T^2\ell_{dS}^2+4M_P^4}}\right)$ & $-8\pi^2M_P^2\ell_{dS}^2\sqrt{\frac{T^2\ell_{dS}^2}{T^2\ell_{dS}^2+4M_P^4}}$\\
     \hline
\end{tabular}
\caption{Comparison between the no boundary, the bubble of something and the spacelike ETW brane example. Importantly, if the ETW brane tension is small, the instanton action of the spacelike ETW brane instanton in the last column becomes small.}
\label{tab_comparison}
\end{table}

In case the underlying theory of quantum gravity allows for ETW branes, further instanton geometries exist. Let us briefly argue that this is a reasonable assumption: First, ETW branes can arise from shrinking extra dimensions \cite{Witten:1981gj}. Second, they appear naturally in type IIA string theory in the form of O8 planes. Finally, they are predicted to exist in any quantum gravity theory by the more general Cobordism Conjecture \cite{McNamara:2019rup}. An early discussion of ETW branes with tension appears in \cite{Jourjine:1983du}. Recent studies of ETW brane examples include \cite{GarciaEtxebarria:2020xsr, Buratti:2021fiv, Angius:2022aeq, Angius:2022mgh, Blumenhagen:2022mqw, Blumenhagen:2023abk, Hassfeld:2023tid, Huertas:2023syg, Angius:2023uqk,Blanco-Pillado:2023aom,Blanco-Pillado:2023hog,Sugimoto:2023oul}.\footnote{We note in particular that, using an explicit regularization of the 4d ETW brane tension, actions can be obtained in a universal way from the 4d EFT\cite{ Hassfeld:2023tid,Sugimoto:2023oul}.}

Once we allow for ETW branes, a further creation process becomes possible as illustrated in the second column of Table~\ref{tab_comparison}. This goes back to \cite{Hawking:1998bn, Turok:1998he}, with the interpretation in terms of ETW branes provided by \cite{Garriga:1998ri,Bousso:1998pk,Blanco-Pillado:2011fcm}. Following our recent study of this process in  \cite{Hassfeld:2023tid}, we advocate the name `bubbles of something', since this is the exact counterpart of vacuum decays due to `bubbles of nothing'. As seen in the table, the instanton contains a spherical ETW brane, characterized by a tension $T$. The latter relates to the extrinsic curvature scalar $\mathcal{K}$ as
\begin{align}
    M_P^2\mathcal{K}=\frac{3}{2}T\,.\label{K_T}
\end{align}
In the considered dS model, this determines the critical bubble radius to be\footnote{
Note 
that, if the ETW brane tension is positive, bubbles of something can also lead to the creation of Minkowski or Anti-de Sitter (AdS) universes. In particular, given that one expects the 10d type-IIB theory to possess a non-SUSY and hence presumably positive-tension ETW brane \cite{McNamara:2019rup}, a corresponding ball-shaped universe with eternally expanding spherical boundary can be created.
}
\begin{align}
    R=\frac{2M_P^2\ell_{dS}}{\sqrt{T^2\ell_{dS}^2+4M_P^4}}\,.\label{R_T}
\end{align}
If $T>0$, the instanton geometry is a patch of a 4-sphere which is smaller than a hemisphere and bounded by an ETW brane.
If $T<0$, the patch is larger than a hemisphere.
As before, the euclidean instanton can be `cut in half' and glued to its Lorentzian analytic continuation, which describes an expanding universe with ETW brane boundary. The instanton action is given by
\begin{align}
    \mathcal{S}_{bos} = -4\pi^2M_P^2\ell_{dS}^2\left(1-\frac{T\ell_{dS}}{\sqrt{T^2\ell_{dS}^2+4M_P^4}}\right)\,.\label{S_BoS}
\end{align}
We see that $|\mathcal{S}_{bos}|$ becomes small for $T\to +\infty$ and approaches the no-boundary action $\mathcal{S}_{nb}=-8\pi^2M_P^2\ell_{dS}^2$ for $T\to -\infty$.  The FLRW universes created in this scenario have negative spatial curvature, similarly to those in creation process of \cite{Cespedes:2023jdk}.

Now we come to our new proposal, the geometry in the right column of Table \ref{tab_comparison}, relying again on ETW branes.  Here, the instanton is a portion of the 4-sphere representing euclidean dS. It is bounded by two ETW branes, each having $S^3$ geometry and radius $R$.  As before, the extrinsic curvature and the tension are related through Eq.~\eqref{K_T} making it necessary for the tension to be negative.
The ETW brane radius is again determined by \eqref{R_T}. This instanton can then be cut at the equator and glued to standard Lorentzian de Sitter space.

We note that, locally, i.e.~focusing on the vicinity of the equator of the sphere and of the waist of the de Sitter hyperboloid, this corresponds to the standard procedure of gluing the euclidean geometry to its analytic continuation.
Globally, this is not the case since there is a boundary in the `past' of the euclidean region and no corresponding deviation from a pure de Sitter space in the Lorentzian geometry. One could in principle improve on this by including a field, the profile of which signals the proximity of the ETW brane, and analytically continuing such a more complete instanton solution. However, we believe that our approach is sufficiently precise under the assumption that any such field is stabilized away from the ETW brane, such that the `local' analytic continuation described above can be trusted.
 
Thus, following the Coleman-de Luccia procedure and analytically continuing the instanton to Lorentzian signature, the nucleated on-shell space does not contain an ETW boundary, cf.~the right column of Table \ref{tab_comparison}.
The ETW brane we used extends in the euclidean direction and marks the `beginning of time', similarly to what is discussed in \cite{McGreevy:2005ci,Angius:2022mgh}.
Our spacelike ETW brane appears only in the off-shell region. To convince readers who doubt that spacelike ETW branes exist at all, we recall the situation in the more established Witten bubble-of-nothing \cite{Witten:1981gj} nucleation process: There, the locus where the $S^1$ shrinks effectively represents an ETW brane and part of it, in the of-shell region, must be spacelike for the decay to occur. 

The instanton action for our new creation process receives contributions from two boundaries and the bulk volume in between. It reads
\begin{align}
    \mathcal{S}_b=-8\pi^2M_P^2\ell_{dS}^2\sqrt{\frac{T^2\ell_{dS}^2}{T^2\ell_{dS}^2+4M_P^4}}\label{S_SL_ETW}
\end{align}
and vanishes if the ETW brane tension approaches zero. This is intuitively clear if one observes from \eqref{R_T} that, in the limit of vanishing tension, the radius of the ETW branes approaches $\ell_{dS}$, both branes approach the equator, and hence the instanton volume goes to zero.
We provide a path-integral saddle-point derivation of the purely geometric result \eqref{S_SL_ETW} in Sect.~\ref{sect_path_integral}.

Considering the lower image in the right column of Table~\ref{tab_comparison} and recalling Eq.~\eqref{R_T}, we see that the limit $T\to 0_-$ corresponds to the ETW brane approaching the waist of the hyperboloid from below. One may then speculate what happens to this geometry if the tension is further raised, becoming small and positive. The natural expectation is a `beginning of the universe' on a boundary in the on-shell regime. This boundary would have to cut the future dS hyperboloid just above its waist. We would now be dealing with an on-shell, lorentzian version of the ETW brane. The new boundary would relate to our previously discussed euclidean ETW brane in the same way in which the on-shell S-brane \cite{Gutperle:2002ai} (see also \cite{Hull:1998vg})
relates to the off-shell euclidean brane of stringy instantons \cite{Becker:1995kb}. We do not follow this speculation further, leaving a proper study to future work.

\begin{figure}
	\centering
	\includegraphics[width=0.25\linewidth]{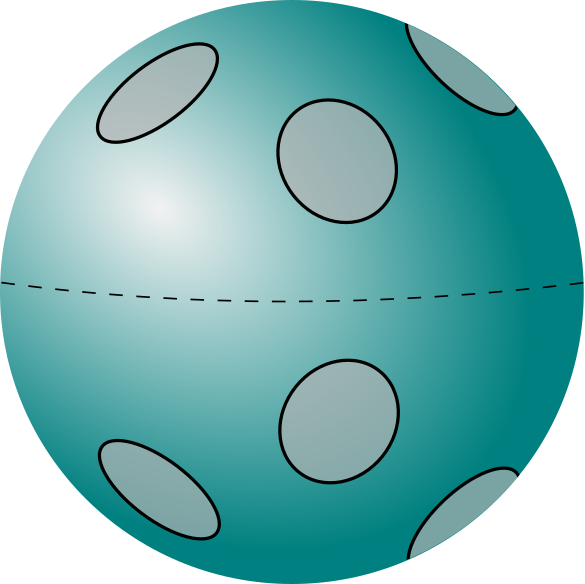}
	\caption{Illustration of the `Swiss Cheese Instanton'. 
 }%
	\label{fig_Swiss_Cheese}
\end{figure}

It is clear that the existence of ETW branes allows for further instanton geometries.  For example, one could 
consider spheres with more than two holes cut out, respecting reflection symmetry along the euclidean time axis and with none of the holes intersecting the equator. A corresponding instanton is illustrated in Fig.~\ref{fig_Swiss_Cheese}.
An appropriate name for such a geometry would be `Swiss Cheese Instanton'.
The radius of each hole is again determined by the tension of the ETW brane \eqref{R_T}.
Each hole reduces the value of $|\mathcal{S}_{instanton}|$. 
As a result, when adopting the Linde/Vilenkin choice of sign, the
Swiss Cheese Instanton with maximal number of holes might turn out to be the most likely creation process for a given de Sitter space.\footnote{
A 
similar observation relevant for the Hartle-Hawking choice of sign was made
in \cite{Fischler:1990pk}. There, it was argued that an instanton geometry with multiple spheres glued together through small tubular regions is preferred in comparison to the single sphere of the original proposal.}

\subsection{Explicit path integral analysis of the boundary process}\label{sect_path_integral}
Let us further study the boundary process by following \cite{Coleman:1977py,Callan:1977pt,coleman1980gravitational,hartle1983wave} and study the euclidean gravitational path integral.
It has been proven to be a useful tool for computing vacuum decay rates as well as deriving thermodynamic properties of black holes.

Following e.g.~\cite{Halliwell:1989myn, Wiltshire:1995vk, Lehners:2023yrj}, we consider a minisuperspace setup with metric 
\begin{align}
    ds^2=\frac{N^2}{q(t)}dt^2+q(t)d\Omega_3^2\,,\label{metric_PI}
\end{align}
a time interval $t\in[0,1]$ and a gauge where $N$ is constant. Setting $M_P=1$, 
the action reads
\begin{align}
    S[q,N]=2\pi^2\int_0^1 dt\left(-\frac{3\dot q^2}{4N}+N(\Lambda q-3)\right)+2\pi^2q_0^{3/2}\frac{N}{|N|}T\,.\label{action_PI}
\end{align} 
The first term is nothing but the Einstein-Hilbert action evaluated for the metric \eqref{metric_PI}.
Since our geometry has two boundaries, defined by $q(0)=q_0$ and $q(1)=q_1$, a standard GHY term is used in the evaluation. The second term in \eqref{action_PI} is the contribution of an ETW brane with tension $T$ located at the $q=q_0$ boundary. Its form can be understood as follows: A domain wall with tension $T$ at a fixed physical position makes the following contribution to the action
\begin{align}
 \int d^4x \sqrt{g}\,\delta\left(\frac{N}{\sqrt{q}}(t-t_0)\right)T=\int d^4x Nq\delta(t-t_0)\left|\frac{\sqrt{q}}{N}\right|T=2\pi^2q(t_0)^{3/2}\frac{N}{|N|}T\,.
\end{align}
Setting $t_0=0$, the volume of the spherical brane eventually becomes $2\pi^2q_0^{3/2}$ and \eqref{action_PI} follows.

We are interested in the amplitude
\begin{align}
G[q_1;q_0]=\int_{q(0)=q_0}^{q(1)=q_1}\mathcal{D}qdN \,e^{-S[q,N]}\label{amplitude_def}
\end{align}
for an initial three-sphere of radius $q_0^{1/2}$ to evolve to a sphere with radius $q_1^{1/2}$. The wave function of the universe follows after integrating over all possible radii of the ETW brane: 
\begin{align}
    \Psi(q_1)=\int dq_0\, G[q_1,q_0]\,.\label{Psi_boundary_def}
\end{align}
We will evaluate it in the saddle point approximation.
The saddle points of $G[q_1,q_0]$ are determined by solutions of the equation of motion and the Hamiltonian constraint:
\begin{align}
    \ddot{q}=-\frac{2}{\ell_{dS}^2}N^2\,,\qquad -\frac{3}{4N^2}\dot{q}^2+3=\frac{3q}{\ell_{dS}^2}+2T\delta(N)q_0^{3/2}\,.\label{eom_and_constraint}
\end{align}
Using Dirichlet boundary conditions, as specified above, solutions of the former equation read
\begin{align}
    \bar{q}(t)=-\frac{N^2t^2}{\ell_{dS}^2}+\left(\frac{N^2}{\ell_{dS}^2}+q_1-q_0\right)t+q_0\,.
\end{align}
There are four values of $N$ ensuring that $\bar q$ is also a solution to the Hamiltonian constraint:
\begin{align}
    N=\pm \ell_{dS}^2\left[\left(1-\frac{ q_0}{\ell_{dS}^2}\right)^{1/2}\pm \left(1-\frac{q_1}{\ell_{dS}^2}\right)^{1/2}\right]\,,
\end{align}
which for $q_0,q_1<\ell_{dS}^2$ are real-valued. Following the intuition that a small sphere nucleates and grows, eventually going on-shell, we restrict attention to $q_0<q_1$.
As discussed e.g.~in \cite{Halliwell:1989myn,Wiltshire:1995vk}, in the case without ETW brane, the four solutions define saddle points potentially contributing to the path integral. 
There are various proposals for which saddle points are relevant.
The two solutions with smaller $N^2$ describe geometries fitting within one hemisphere.
By contrast, the solutions with larger $N^2$ corresponds to geometries containing the equator. 
Hartle and Hawking argue that the saddle point with smaller $N^2$ and $N>0$ defines the wavefunction, meanwhile the Linde/Vilenkin proposal corresponds to choosing the saddle point with the same $N^2$ but $N<0$.
Following these arguments, the amplitude calculated in the saddle point approximation is given by
\begin{align}
G[q_1;q_0]\propto\exp(\pm\left[4\pi^2\ell_{dS}^2\left(1-\frac{q_0}{\ell_{dS}^2}\right)^{3/2}-2\pi^2q_0^{3/2}T-4\pi^2\ell_{dS}^2\left(1-\frac{q_1}{\ell_{dS}^2}\right)^{3/2}\right])\,,\label{Amplitude_PI}
\end{align}
with the positive and negative sign corresponding to the Hartle-Hawking and Linde/Vilenkin choice respectively.
To find $\Psi$ from \eqref{Psi_boundary_def}, we perform the integral over $q_0$, again in a saddle point approximation. 
Extremizing \eqref{Amplitude_PI} with respect to $q_0$ yields the condition $T<0$ and
\begin{align}
    q_0=\frac{4\ell_{dS}^2}{T^2\ell_{dS}^2+4}\,,
\end{align}
which is precisely the result \eqref{R_T}.
Ultimately, we find for the wavefunction
\begin{align}
\Psi(q_1)\propto\exp(\pm\left[4\pi^2M_P^2\ell_{dS}^2\sqrt{\frac{T^2\ell_{dS}^2}{T^2\ell_{dS}^2+4M_P^4}}-4\pi^2\ell_{dS}^2\left(1-\frac{q_1}{\ell_{dS}^2}\right)^{3/2}\right])\,.
\label{esol}
\end{align}
As explained in \cite{Halliwell:1989myn,Wiltshire:1995vk}, one obtains approximate solutions to the WDW equation in the regime $q_1>\ell_{dS}^2$ using the WKB matching procedure. Specifically, we analytically continue \eqref{esol} to the on-shell regime and enforce reality in the Hartle-Hawking case or the outgoing-wave condition in the Linde-Vilenkin case:
\begin{align}
    \Psi_{\rm HH}(q_1)&\propto \exp(4\pi^2M_P^2\ell_{dS}^2\sqrt{\frac{T^2\ell_{dS}^2}{T^2\ell_{dS}^2+4M_P^4}}\,)\cos(4\pi^2\ell_{dS}^2\left(\frac{q_1}{\ell_{dS}^2}-1\right)^{3/2}-\frac{\pi}{4})\,,\label{Psi_HaHa}\\
    \Psi_{\rm LV}(q_1)&\propto \exp(-4\pi^2M_P^2\ell_{dS}^2\sqrt{\frac{T^2\ell_{dS}^2}{T^2\ell_{dS}^2+4M_P^4}}-i4\pi^2\ell_{dS}^2\left(\frac{q_1}{\ell_{dS}^2}-1\right)^{3/2})\,.\label{Psi_LV}
\end{align}
For an analysis of the Swiss-Cheese instanton, one would need to leave the minisuperspace approximation, which is beyond the scope of this note.

\subsection{Creation rates}
Given the various instanton geometries from Sect.~\ref{sect_instantons} or alternatively, the wavefunctions computed in Sect.~\ref{sect_path_integral}, one may compute the vacuum creation rates.
For vacuum decays, the instanton action is related to the decay rate $\Gamma$ via \cite{coleman1980gravitational}
\begin{align}
    \Gamma\propto\exp(-B)\,,\qquad B=\mathcal{S}_{instanton}-\mathcal{S}_{vacuum}\,,
\end{align}
with $\mathcal{S}_{vacuum}$ the action of the decaying vacuum.
The decay rate is then interpreted as decays per time and volume in the parent vacuum.
For the creation from nothing, there does not exist a parent vacuum such that one may define \cite{Vilenkin:1982de,Vilenkin:1983xq,hartle1983wave}
\begin{align}
    \Gamma\propto \exp(-\mathcal{S}_{instanton})\,,
\end{align}
without having the interpretation of creations per time and volume. 
Rather, physical statements are relative creation rates of the form
\begin{align}
    \Gamma_i/\Gamma_j\,,
\end{align}
for different creation processes $i$ and $j$.
Then, noting that the instanton action is always negative, one obtains 
\begin{align}
    \Gamma\propto \exp(+|\mathcal{S}_{instanton}|)\,,\label{rate_HaHa}
\end{align}
with $|\mathcal{S}|$ growing for larger $\ell_{dS}$ (and fixed $T$) for all discussed instanton geometries. Maybe counter-intuitively, this implies that it is simpler to quantum mechanically nucleate a large rather than small dS sphere. The origin of this behaviour can be traced to the negative sign in front of the kinetic term of the conformal mode of gravity.

By contrast, relying on standard tunneling behaviour, Linde \cite{Linde:1983mx} and Vilenkin \cite{Vilenkin:1984wp} suggested that the instanton action should relate to the creation rate according to
\begin{align}
\Gamma\propto\exp(+\mathcal{S}_{instanton})=\exp(-|\mathcal{S}_{instanton}|)\,,\label{rate_LV}
\end{align}
thus favoring small universes.  It remains an open question whether the Hartle-Hawking \eqref{rate_HaHa} or Linde/Vilenkin \eqref{rate_LV} sign choice is the right one in quantum gravity.

Alternatively, one may derive the vacuum creation rates from the wavefunctions \eqref{Psi_HaHa},\eqref{Psi_LV}. Following \cite{Halliwell:1989myn,Wiltshire:1995vk}, one finds that the creation rates agree with \eqref{rate_HaHa} and \eqref{rate_LV} for the Hartle-Hawking and Linde/Vilenkin wavefunction respectively.

Let us compare the creation rates of the three instanton geometries of Table \ref{tab_comparison}.
It is clear that
\begin{align}
|\mathcal{S}_{nb}|>|\mathcal{S}_{bos}|\quad\text{ and }\quad|\mathcal{S}_{nb}|>|\mathcal{S}_{b}|\label{S_ineq}
\end{align}
always hold. Hence for the Hartle-Hawking choice of sign the no-boundary process always dominates the creation of de Sitter universes.\footnote{The bubble of something may still be relevant since it also occurs for positive tension ETW branes and allows for the creation of Minkowski and AdS vacua, which are not accessible in the no-boundary process.
For example, the ETW brane bounding an empty 10d type IIB string theory universe conjectured to exist in \cite{McNamara:2019rup} breaks supersymmetry and hence presumably has positive tension. As a result, it may be responsible for the creation of plain type IIB universes.}

The situation is different when adopting the Linde/Vilenkin choice of sign. 
Here, due to \eqref{S_ineq}, a bubble of something or a boundary instanton are the most likely processes. 
Furthermore, using negative-tension branes only, it is straightforward to see that $|\mathcal{S}_{bos}|>|\mathcal{S}_{b}|$ such that a boundary process wins. 
This becomes particularly clear for small brane tensions, i.e. $|T|\ell_{dS}\ll M_P^2$. Then the instanton actions take the form
\begin{align}
    |\mathcal{S}_{nb}|&=8\pi^2M_P^2\ell_{dS}^2\,,\\
    |\mathcal{S}_{bos}|&\simeq 4\pi^2M_P^2\ell_{dS}^2\,,\\
    |\mathcal{S}_{b}|&\simeq 4\pi^2|T|\ell_{dS}^3\ll 4\pi^2M_P^2\ell_{dS}^2\,.
\end{align}
Hence, for small negative ETW brane tensions, the no-boundary action is of the same order as the bubble of something action while the boundary action is parametrically smaller. 

The bubble of something becomes particularly relevant for positive tension ETW branes, when the boundary instanton does not exist: In the regime $T\ell_{dS}\gg M_P^2$ we have
\begin{align}
    |\mathcal{S}_{nb}|=8\pi^2M_P^2\ell_{dS}^2\gg \frac{8\pi^2M_P^6}{T^2} \simeq |\mathcal{S}_{bos}|\,.
\end{align}

The tension of a brane is clearly similar to a $3d$ vacuum energy.
As is well known, it proves difficult to realize string theory vacua with large positive vacuum energy. 
As a result, one may speculate that ETW branes with large positive tension are also problematic.  On the other hand, string-theoretic examples of ETW branes with negative tension exist \cite{Witten:1981gj, Klebanov:2000hb, GarciaEtxebarria:2020xsr}.\footnote{
The small negative tension of the ETW brane of \cite{GarciaEtxebarria:2020xsr} becomes apparent using the 4d EFT approach as developed in \cite{Hassfeld:2023tid}.
}
As a consequence, the creation of vacua via the boundary process may turn out to be dominant in the string landscape.

One may study a geometrical model for the ETW brane required to realize our boundary process. The simplest such model is presumably the shrinking $S^1$ of Witten's bubble of nothing~\cite{Witten:1981gj}. To implement this, consider a universe with spatial topology $S^n\times S^1$. If the $S^1$ radius is small and the $S^n$ radius is large, one may interpret the spacelike part of the universe as effectively $n$-dimensional.  In this setting, our boundary instanton corresponds to a creation process in which the $S^1$ direction shrinks to zero size in the past while the $S^n$ remains finite.

Intriguingly, this process has been studied in \cite{Conti:2014uda,Fanaras:2021awm,Fanaras:2022twv} for the so called Kantowski-Sachs model, where the universe has the topology of $S^2\times S^1$.
Both the Hartle-Hawking no-boundary and the Linde/Vilenkin tunneling initial conditions were analyzed.  For us, the tunneling initial conditions are most interesting since this is where we expect that the boundary instanton could dominate. For this case, it was shown \cite{Fanaras:2022twv} that in the limit of a small $S^1$, universes with a large $S^2$ radius are favored to nucleate. One may hence interpret the Picard-Lefschetz-based results of \cite{Fanaras:2022twv} as supporting our suggestion of an (in this case effective) boundary creation process.

\section{Creation of toric universes\\ with zero cosmological constant} 
As has been noticed long ago \cite{Zeldovich:1984vk,Coule:1999wg,Linde:2004nz}, a toric universe with positive cosmological constant can classically evolve from zero radius without the need of an off-shell region. A more complicated analysis is required to include quantum effects, such as fluctuations of additional fields or the Casimir energy in the calculation of the creation rate.

Such a creation from nothing process is, of course, also conceivable if the cosmological constant of the torus vanishes. But in this case the created universe would classically not expand and it can only grow due to quantum fluctuations.

\begin{figure}
	\centering
	\includegraphics[width=0.25\linewidth]{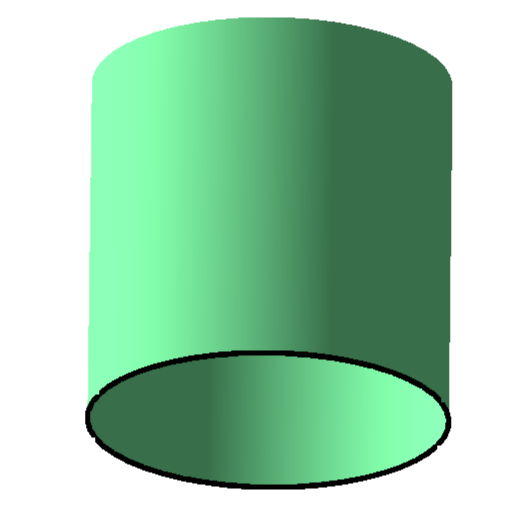}
	\caption{The spacetime geometry of a toric universe nucleating from nothing.}%
	\label{fig_Torus_Instanton}
\end{figure}

One can, however, use a variant of our boundary instanton with an ETW brane with vanishing tension to create a finite-size toric universe with zero cosmological constant from nothing. Unlike dS universes, the so created spacetime could even be supersymmetric, so this could for example apply to M-theory with its familiar ETW brane or to type-IIA string theory with an O8/D8 tensionless boundary. The process is illustrated for the 1+1 dimensional case in Fig.~\ref{fig_Torus_Instanton}, but it works analogously in any number of dimensions. 

The instanton, in the classical sense of a bounce solution, would be just $T^{d-1}\times I$, with $I$ an interval of euclidean time of arbitrary length. At the two boundaries, corresponding to the two ends of $I$, there are two ETW branes with the geometry of $T^d$. The action clearly vanishes. We can cut this instanton in the middle and glue it to the future half of a static, Lorentzian, flat torus universe.
Note that we can choose $I$ arbitrarily small since no off-shell region is really required. 
There is also no region where the size of the torus is small, such that the issues of corrections at small torus volume do not arise.  We admit that it is rather counter-intuitive that this process is in no way suppressed if the torus volume becomes large. However, we were not able to find a hard argument against this conclusion at the present level of precision of our analysis.

Clearly, if we were to wrap our torus by a conventional brane or domain wall rather than by an ETW brane, our conclusion would be completely standard: The instanton would be just a standard brane instanton (see e.g.~\cite{Blumenhagen:2009qh}). Its action would be the product of torus volume and brane tension, going to zero if the tension vanishes. For an ETW brane, the situation is different since its action contains a tension term, just like for a usual brane, but in addition the Gibbons-Hawking-York term, governed by the Planck scale and potentially introducing a wrong-sign kinetic term, as for the gravity conformal mode. Nevertheless, wrapping this brane on a torus gives, in the limit of zero tension, a zero-action instanton. 

While we believe that this puzzling result deserves further study, it naively appears that zero-tension ETW branes are able to nucleate (and also terminate) toric universes without any action cost. Torus universes would then undergo a continuous production and destruction cycle. 

\section{Conclusions}
Famously, a de Sitter universe can be created from nothing using the smooth geometry of the No-Boundary Proposal. If the relevant theory of quantum gravity allows for end-of-the world (ETW) branes, two further options exist: One nucleates a ball of dS space with an ETW-brane boundary from nothing. The other, being the main subject of this paper, nucleates a compact, boundary-free sphere but, in contrast to the No-Boundary Proposal, with a spacelike ETW brane at the beginning of the universe. One may think of this as the `Boundary Proposal'.

The Boundary Proposal requires the ETW brane to have negative tension. Explicitly, its instanton action approaches that of the No-Boundary Proposal if the tension goes to minus infinity and it vanishes if the tension approaches zero. If one uses the Linde/Vilenkin (or tunneling) sign choice to compare the rates of creation processes, the Boundary Proposal is dominant if ETW branes with negative or zero tension exist.

A maybe surprising implication of our proposal is that, in the presence of zero-tension ETW branes, torus universes with vanishing cosmological constant can be created from nothing or decay without any exponential suppression or enhancement. Indeed, a flat ETW brane cutting a euclidean torus universe across the time-direction does not contribute to the action. It will be important to understand whether and how this puzzling result changes if fluctuations of the ETW brane are included. Maybe this will require developing a better understanding of the relation between time- and spacelike ETW branes. Moreover, a deeper understanding can hopefully be achieved by investigating carefully various geometrical models for spacelike ETW branes, starting with the simplest case of the effective ETW brane in Witten's bubble of nothing.

Finally, we recall that the No-Boundary Proposal with Hartle-Hawking sign choice famously has the phenomenological problem of predicting the lowest possible starting point for the inflaton $\phi$ on its potential plateau. This has very recently been emphasized in \cite{Maldacena:2024uhs}.
At first sight, one could think that our boundary proposal can help since the inflaton can be forced to take a certain fixed value at the ETW-brane, in the sense of a Dirichlet-boundary condition for $\phi$. If this value lies high enough on the plateau, correct predictions could follow. However, for the Hartle-Hawking sign-choice our boundary creation process is subleading w.r.t.~the no-boundary process, so in total we still get the wrong cosmological prediction.  For the Linde-Vilenkin sign choice, our ETW brane proposal can be dominant and can be used to fix the initial value of $\phi$.

Very recently, a new creation process has been proposed in \cite{Betzios:2024oli} which has interesting similarities to our `boundary process'. The idea is that the past of our compact universe is a euclidean wormhole leading to euclidean AdS, with the inflaton being the field which interpolates between our inflationary plateau and the AdS minimum defining the euclidean past. One may say that \cite{Betzios:2024oli} abandons the no-boundary creation process in favor of postulating a spacelike domain wall from euclidean dS to euclidean AdS space in the past.\footnote{It would be interesting to study possible relations to the domain walls to euclidean AdS of \cite{Hertog:2011ky,Hartle:2012tv}.} In our proposal, the place of this spacelike domain wall is, in a sense, taken by a spacelike ETW brane. However, since we still work in the general framework of creation from nothing, the no-boundary process is also there and is more enhanced by its action. In principle, this might also be a concern for the proposal of \cite{Betzios:2024oli}, given in particular that the positive action of euclidean AdS tends to suppress any process in which it appears. Hence, in the case of Hartle-Hawking sign choice, it seems necessary to forbid the creation of small spheres for either the domain wall or ETW brane to dominate the beginning of the universe. It would clearly be important to establish a more detailed relation between No-Boundary, Boundary and further, alternative creation processes.

\section*{Acknowledgments}
We thank I\~{n}aki Garc\`{i}a Etxebarria and Daniel Schiller for valuable discussions and comments. This work was supported by the Deutsche Forschungsgemeinschaft (DFG, German Research Foundation) under Germany’s Excellence Strategy EXC 2181/1 - 390900948 (the Heidelberg STRUCTURES Excellence Cluster).

\bibliographystyle{utphys}
\bibliography{References_Spacelike_ETW_branes}
\end{document}